\documentclass{aa}
\usepackage{graphicx}
\usepackage{txfonts}
\begin{document}
   \title{AcDc - A new code for the NLTE spectral analysis of accretion
   discs: application to the helium CV AM\,CVn}

   \author{T. Nagel\inst{1}
          \and
          S. Dreizler\inst{1,2}
          \and
          T. Rauch\inst{1,3}
          \and
          K. Werner\inst{1}
          }

   \offprints{T. Nagel}

   \institute{Institut f\"ur Astronomie und Astrophysik, Abteilung
     Astronomie, Universit\"at T\"ubingen, Sand 1, 72076 T\"ubingen, Germany\\
     \email{nagel@astro.uni-tuebingen.de}
     \and
     Universit\"atssternwarte G\"ottingen, Geismarlandstr. 11, 37083 G\"ottingen, Germany
     \and
     Dr.-Remeis-Sternwarte, Universit\"at Erlangen-N\"urnberg, Sternwartstr. 7,
     96049 Bamberg, Germany 
     }

   \date{Received xx; accepted xx}

   \abstract{ We present a recently developed code for detailed NLTE
   calculations of accretion disc spectra of cataclysmic variables and
   compact X-ray binaries. Assuming a radial structure of a standard
   $\alpha$-disc, the disc is divided into concentric rings. For each disc
   ring the solution of the radiation transfer equation and the structure
   equations, comprising the hydrostatic and radiative equilibrium, the
   population of the atomic levels as well as charge and particle
   conservation, is done self-consistently.
   Metal-line blanketing and irradiation by the central object are taken
   into account. As a first application, we show the influence of different
   disc parameters on the disc spectrum for the helium cataclysmic variable
   AM\,CVn.
   
   }
   \authorrunning{T. Nagel et al.}
   \titlerunning{AcDc - A new code for the NLTE spectral analysis of
   accretion discs}
   \maketitle

\section{Introduction}
Accretion discs are components of objects as diverse as proto-planetary
systems, active galactic nuclei, cataclysmic variables or X-ray binaries.
A high fraction of the luminosity of these systems may be generated by
the accretion disc itself. 
To understand these objects and interpret the observational data a
model of the accretion disc as reliable as possible is 
necessary. The aim of our work was the development of a 
program package for the calculation of synthetic spectra and vertical structures
of accretion discs considering the physical processes in the disc as
accurately as possible.  

A fully three-dimensional radiation-hydrodynamic treatment is
presently still impossible because of the enormous numerical costs. In the case of a 
geometrically thin $\alpha$-disc (Shakura \& Sunyaev 1973), where the disc thickness 
is significantly smaller than the disc diameter, the radial and vertical
structures can be decoupled.  Under the assumption of axial symmetry and  
by dividing the disc into concentric rings the determination of
the vertical structure becomes a one-dimensional problem. The dissipated
energy in each disc ring is radiated away at the disc surface, the energy
flux can be expressed as effective temperature. As an approximation, it can
be assumed that the disc rings are radiating like black bodies. An
improvement of the models can be obtained by describing the disc rings by
stellar atmosphere models of the same effective temperature, see
e.g. Kiplinger (1979), Mayo et al. (1980) or La Dous (1989) in the case of
cataclysmic variables  and Kolykhalov \& Sunyaev (1984) or Sun \& Malkan
(1989) in the case of AGN. Unfortunately, neither black bodies nor stellar
atmosphere models reproduce spectra of accretion discs in an adequate
manner (Wade 1988). Meyer \& Meyer-Hofmeister (1982), Cannizzo \& Wheeler
(1984) and Cannizzo \& Cameron (1988) calculated the radiative transfer
using the diffusion assumption. This assumption is only valid at large
optical depths, but the spectrum is generated at optical depths around
$\tau\sim 1$, where neither the diffusion assumption nor the assumption of
local thermodynamic equilibrium (LTE) is fulfilled. Only solving the
radiative transfer equation self-consistently with the structure equations
allows the calculation of realistic accretion disc spectra (Kriz \& Hubeny
1986, Shaviv \& Wehrse 1986, Stoerzer et al. 1994). In the last
decade much work in this field was done e.g. by Hubeny \& Hubeny (1997,
1998) and Hubeny et al. (2000, 2001).  

Following the path mentioned above, we have developed our program package {\sc
AcDc} ({\sc Ac}cretion {\sc D}isc {\sc c}ode) for the detailed
calculation of vertical structures and NLTE spectra of accretion
discs. For each disc ring the equations of radiative and hydrostatic
equilibrium as well as the NLTE rate equations for the population numbers
of the atomic levels are solved consistently with the radiation transfer
equation under the constraint of particle number and charge
conservation. Full metal-line blanketing as well as irradiation of the
accretion disc by the central object can be considered. By integrating the
spectra of the individual disc rings, one obtains a complete disc
spectrum for different inclination angles, where the spectral lines are
Doppler shifted according to the radial component of the Kepler rotation.  This
is shown in Sect. 2, whereas in Sect. 3 we show first applications of
the developed program package to examine the influence of different
parameters on the vertical structure and the spectrum of an accretion disc
model for the helium cataclysmic variable AM\,CVn.

\section{Vertical Structure of Accretion Discs}
\begin{figure}
\centering
\includegraphics[height=7cm]{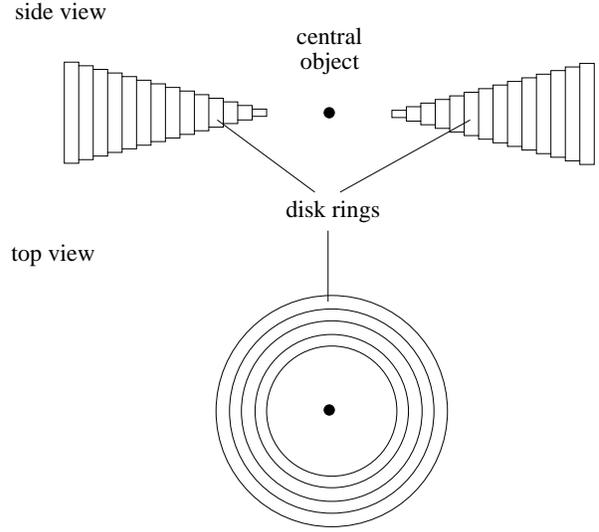}
\caption{\label{ringe}Geometry of the accretion disc, divided into concentric rings.}
\end{figure}

We assume a geometrically thin (disc thickness is significantly smaller than the disc
diameter) stationary accretion disc ($\alpha$-disc, Shakura \& Sunyaev
1973). We also assume that the mass of the disc is much smaller than the mass of the
central object, so we can neglect self-gravitation. Introducing the surface
mass density $\Sigma$ as
\begin{equation}
\Sigma\,=\,2\,\int\limits_{0}^{H/2}\rho\,dz\,,
\end{equation}
with mass density $\rho$, geometrical height $z$ above the midplane and
total disc height $H$, the radial dependence of $\Sigma$ reads following
Shakura \& Sunyaev (1973)
\begin{equation}
\nu\,\Sigma(R)\,=\,\frac{\dot{M}}{3\pi}\left(1\,-\,\left(\frac{R_\star}{R}\right)^{1/2} \right)\,.
\end{equation}
Here, $R$ denotes the distance from the central object, $R_\star$ the
radius of the central object, $\dot{M}$ the mass accretion rate and $\nu$
the kinematic viscosity, defined as 
\begin{equation}
\nu\,=\,\alpha\,c_{\rm{s}}\,H\,,
\label{alpha}
\end{equation}
with sound speed $c_{\rm{s}}$ and the parameter $\alpha$ being a measure
of the efficiency of angular momentum transport through the disc.
The radial distribution of the effective temperature $T_{\rm eff}$ then,
following Shakura \& Sunyaev (1973), can be described by   
\begin{equation}
T_{\rm eff}(R)\,=\,\left[\frac{3GM_\star\dot{M}}{8\pi\sigma
    R^3}\left(1\,-\,\left(\frac{R_\star}{R}\right)^{1/2} \right)
\right]^{1/4}
\label{tglg}
\end{equation}
with $M_\star$ denoting the mass of the central object, $G$ the
gravitational constant and $\sigma$ the Stefan-Boltzmann constant.
The accretion disc is divided into a set of concentric disc rings
(cf. Fig. \ref{ringe}). For each ring the vertical structure is calculated
by solving the set of equations described in the following two subsections,
assuming a plane-parallel geometry.  

\subsection{Radiation Transfer Equation}
\begin{figure}
\centering
\includegraphics[width=0.48\textwidth]{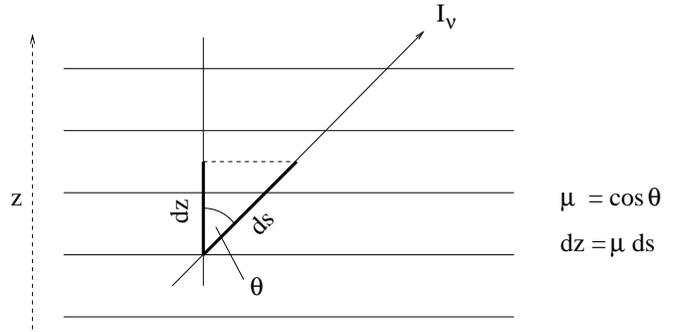}
\caption{\label{rt}Radiation transfer through the accretion
  disc layers.}
\end{figure}

In order to compute the radiation field, which determines the atomic population
numbers, the radiation transfer equation has to be solved. This equation
describes the modification of the specific intensity $I_{\nu}$ of a ray due
to absorption or emission along its path $ds$ through the accretion disc
(cf. Fig. \ref{rt}).
The radiation transfer equation for the shown geometry then reads
\begin{equation}
\mu\,\frac{\partial\,I_{\nu}(\nu,\mu,z)}{\partial\,z}\,=\,-\chi_{\nu}(\nu,z)\,I_{\nu}(\nu,\mu,z)\,+\,\eta_{\nu}(\nu,z)\,.
\end{equation}
with the absorption coefficient $\chi_{\nu}$ and the emission coefficient
$\eta_{\nu}$. Introducing the source function $S_{\nu}$ 
\begin{equation}
  S_{\nu}\,=\,\frac{\eta_{\nu}}{\chi_{\nu}}\,
\end{equation}
the radiation transfer equation is
\begin{equation}
\mu\,\frac{\partial\,I_{\nu}(\nu,\mu,z)}{\partial\,z}\,=\,-\chi_{\nu}(\nu,z)\,(I_{\nu}(\nu,\mu,z)\,-\,S_{\nu}(\nu,\mu,z))\,.
\end{equation}
The solution of this equation is a complicated problem, because
the source function depends on the radiation field itself. Within an
iteration scheme (see below), one solves the radiation transfer 
equation formally assuming the source function known (cf. Mihalas
1978). In our work this is done by a short characteristics
method (Olson \& Kunasz 1987). 
Irradiation of the disc by an external source with a given spectrum is
accounted for by an appropriate boundary condition.

\subsection{Structure Equations}
In order to obtain the atomic population numbers in NLTE all processes populating or
de-populating an atomic level are considered. These are ionisation,
recombination, excitation and de-excitation, caused either by radiation or
collision. Each level $i$ of each ion has one rate equation describing the
modification of the population density $n_i$ with time $t$:
\begin{equation}
\frac{\partial n_i}{\partial t}\,=\,n_i\sum_{i\neq j}P_{ij}\,-\,\sum_{j\neq i}n_jP_{ji}\,.
\end{equation}
$P_{ij}$ denotes the rate coefficients, consisting of radiative and
collisional components. Since we assume a hydrostatic stratification, we
have  
\begin{equation}
\frac{\partial n_i}{\partial t}\,=\,0\,.
\end{equation}

Assuming that the radial component of the gravitation of the central object
equals the centrifugal force of the Keplerian rotation of the disc, the
hydrostatic equilibrium of the disc only is determined by the vertical
component of the gravitation
\begin{equation}
\frac{dP}{dz}\,=\,-\frac{G\,M_\star}{R^3}\,z\,\rho\,,
\label{hydros}
\end{equation}
with $P$ denoting the total (gas and radiation) pressure.

Another fundamental equation of the vertical structure describes the local
energy balance:
\begin{equation}
E_{\rm{mech}}\,=\,E_{\rm{rad}}\,+\,E_{\rm{conv}}\,.
\end{equation}
The viscously generated energy $E_{\rm{mech}}$ is equal to the radiative
energy loss $E_{\rm{rad}}$; the convective energy $E_{\rm{conv}}$ term is
neglected in the following. For standard $\alpha$-discs, the viscously
generated energy reads 
\begin{equation}
E_{\rm{mech}}\,=\,\nu\Sigma\left(R\frac{d\omega}{dR}\right)^2\,=\,\frac{9}{4}\nu\Sigma\frac{G\,M_\star}{R^3}
\end{equation}
and 
\begin{equation}
E_{\rm{rad}}\,=\,4\,\pi\,\int\limits_{0}^{\infty}(\eta(\nu,z)\,-\,\chi(\nu,z)\,J(\nu,z)\,)\,d\nu\,.
\end{equation}
Introducing now the mass column depth $m$ as
\begin{equation}
m(z)\,=\,\int\limits_{z}^{\infty}\rho\,dz\,
\end{equation}
with the total column mass $M_0$ at the midplane the energy dissipated at
each depth is 
\begin{equation}
E_{\rm{mech}}(m)\,=\,\frac{9}{4}\,\frac{G\,M_\star}{R^3}\,\nu(m)\,\rho\,.
\end{equation}
Here $\nu(m)$ denotes the depth-dependent kinematic viscosity
\begin{equation}
\nu(m)\,=\,a\bar{\nu}(\zeta + 1)\left(\frac{m}{M_0}\right)^\zeta\qquad\mbox{with}\qquad
\zeta > 0 
\end{equation}
with the damping factor $\zeta$, which has been introduced by Kriz \&
  Hubeny (1986) to avoid numerical instabilities at the disc surface. 
$\bar{\nu}$ is the depth-averaged kinematic viscosity with 
\begin{equation}
\bar{\nu}\,=\,\frac{1}{M_0}\int\limits_{0}^{M_0}\nu(m)\,dm\,.
\end{equation}
$\bar{\nu}$ corresponds to $\nu$ in Eq. (\ref{alpha}) and can be determined by 
\begin{equation}
\bar{\nu}\,=\,\frac{Rv_{\phi}}{Re}\,=\,\frac{\sqrt{G\,M_\star\,R}}{Re}
\end{equation}
with $v_{\phi}$ denoting the Keplerian angular velocity and $Re$ the effective
Reynolds number (Lynden-Bell \& Pringle 1974). The prescription of the
viscosity using the Reynolds number has the advantage that no further
assumptions concerning first values of disc height and speed of sound have
to be made, as would be necessary using the
$\alpha$-prescription. Furthermore, it is possible to describe the depth
dependency of the viscosity using the column mass as depth variable. The
solution of the energy balance is obtained with a generalised Uns\"old-Lucy
method (Lucy 1964, Dreizler 2003). 

Finally, the total particle density $N$ consists of the sum of the
population numbers $n$ of the NLTE and LTE levels $l$ of all ions $i$ of all
elements $x$ plus the electron density $n_{\rm{e}}$:
\begin{equation}
N\,=\,\sum_{x=1}^{\rm{element}}\,\sum_{i=1}^{\rm{ion}}\,\left(\sum_{l=1}^{\rm{NLTE}}n_{xil}\,+\,\sum_{l=1}^{\rm{LTE}}n_{xil}^* \right)\,+\,n_{\rm{e}}\,.
\end{equation}
The equation of charge conservation reads
\begin{equation}
n_{\rm{e}}\,=\,\sum_{x=1}^{\rm{element}}\,\sum_{i=1}^{\rm{ion}}\,q(i)\left(\sum_{l=1}^{\rm{NLTE}}n_{xil}\,+\,\sum_{l=1}^{\rm{LTE}}n_{xil}^*\right)\,
\end{equation}
with charge $q(i)$ of the ion $i$.

The solution of the system of equations consisting of the radiation
transfer equation, the equations of energy balance and hydrostatic
equilibrium, the rate equations and the equations of charge and particle
conservation is done in an iterative scheme, the so-called Accelerated
Lambda Iteration (ALI, Werner \& Husfeld 1985; Werner et al. 2003). The
input parameters of our models are mass and radius of
the central object, radius of the disc ring, mass accretion rate and Reynolds
number. Furthermore, the atomic data and an appropriate
frequency grid are specified (see e.g. Rauch \& Deetjen 2003).

\subsection{LTE Start Models}
To avoid numerical instabilities at the beginning of the NLTE
calculations it is necessary to establish suitable start models under the
assumption of LTE. In the following, we summarise the main steps creating
such models, according to Hubeny (1990).  

To determine the atomic LTE population numbers $n$ the Boltzmann equation 
\begin{equation}
\frac{n_i}{n_j}\,=\,\frac{g_i}{g_j}\,e^{-(E_i-E_j)/kT}
\end{equation}
and the Saha equation 
\begin{equation}
\frac{n_{\rm{up}}}{n_{\rm{low}}}\,=\,\frac{2}{n_{\rm{e}}}\left(\frac{2\pi m_{\rm{e}}kT}{h^2}\right)^\frac{3}{2}\frac{g_{\rm{up}}}{g_{\rm{low}}}e^{-(E_{\rm{up}}-E_{\rm{low}})/kT}
\end{equation}
are used instead of the NLTE rate equations.
Here, $g$ denotes the statistical weights, $E$ the excitation or ionisation
energy, $m_{\rm{e}}$ the electron mass and $h$ the Planck constant.
Furthermore, in the case of LTE the source function $S_{\nu}$ equals the Planck
function $B_{\nu}$
\begin{equation}
S_{\nu}\,\equiv\,B_{\nu}\,.
\end{equation}
To get an analytical expression for the vertical temperature structure one
combines the first momentum of the specific intensity and the equation of
the energy balance. Then the equation for the vertical temperature
structure in the case of LTE finally reads 
\begin{equation}
T^4\,=\,\frac{3}{4}T_{\rm eff}^4\left(\tau_{\rm{R}}\left(1-\frac{\tau_{\rm{R}}}{2\tau_{\rm{tot}}}\right)+\frac{1}{\sqrt{3}}+\frac{1}{3\epsilon\tau_{\rm{tot}}}\frac{w}{\bar{w}} \right)\,
\end{equation}
with $\epsilon\,=\,\kappa_{\rm{B}}M_0/\tau_{\rm{tot}}$ and 
\begin{equation}
\kappa_{\rm{B}}\,\,=\,\frac{1}{B}\int\limits_{0}^{\infty}(\kappa_{\nu}/\rho)\,B_{\nu}\,d\nu\,.
\end{equation}

The equation of the hydrostatic equilibrium has the same form as in the
case of NLTE, but can alternatively be transformed into a differential equation
of second order:
\begin{equation}
\frac{d^2P}{dm^2}\,=\,-\frac{c_{\rm{s}}^2}{P}\frac{G\,M_\star}{R^3}\,.
\label{hydros2}
\end{equation}
The solution of this equation is done numerically. The upper boundary
condition reads, following Hubeny (1990), 
\begin{equation}
P(1)\,=\,\frac{m_1c_{\rm{s}}^2}{H_{\rm{g}}}\frac{1}{f(\frac{z-H_{\rm{r}}}{H_{\rm{g}}})}
\end{equation}
with
\begin{equation}
f(x)\,=\,\frac{\sqrt{\pi}}{2}e^{x^2}\,k(x)
\end{equation}
and 
\begin{equation}
k(x)\,=\,\frac{2}{\sqrt{\pi}}\int\limits_{x}^{\infty}e^{-t^2}dt\,.
\end{equation}
Here, $H_{\rm{g}}$ and $H_{\rm{r}}$ denote the gas and radiation
pressure scale height, defined as
\begin{eqnarray}
H_{\rm{g}}\,&=&\,\sqrt{\frac{2c_{\rm{g}}^2}{GM_\star/R^3}}\,,\\
H_{\rm{r}}\,&=&\,\frac{\sigma}{c}T_{\rm eff}^4\kappa_H\frac{GM_\star}{R^3}
\end{eqnarray}
with $c_g$ denoting the sound speed associated with gas pressure
\begin{equation}
c_{\rm{g}}^2\,=\,\frac{P_{\rm{g}}}{\rho}\,.
\end{equation}

\subsection{The Synthetic Spectrum}
\begin{figure}
\centering
\includegraphics[width=0.5\textwidth]{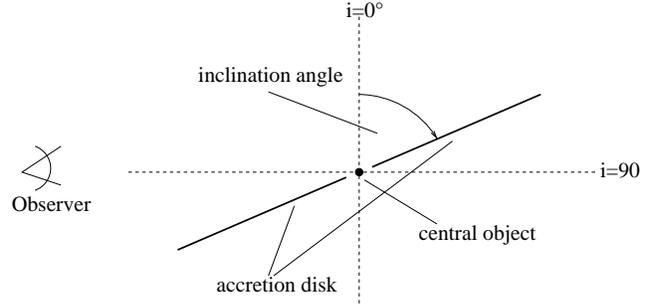}
\caption{Accretion disc geometry.}
\label{Scheibe2}
\end{figure}

Having now calculated the vertical structures and spectra of the individual
disc rings, the ring spectra are integrated to get the spectrum of the
whole accretion disc:
\begin{equation}
I(\nu,i)\,=\,\cos(i)\int\limits_{R_{\rm{i}}}^{R_{\rm{o}}}\int\limits_{0}^{2\pi}I(\nu,i,\phi,r)\,r\,d\phi\,dr\,. 
\end{equation}
Here, $R_{\rm{i}}$ and $R_{\rm{o}}$ denote the inner and outer radius of
the disc, $i$ is the inclination angle (cf. Fig. \ref{Scheibe2}). The
integration over the azimuthal angle $\phi$ is done in intervals of $1^\circ$.
In this last step, spectral lines become broadened due to the Keplerian
rotation of the disc.

\section{Synthetic Spectrum for AM\,CVn}
In order to test and gain experience with {\sc AcDc} our first application
was the calculation of the synthetic spectrum of AM\,CVn for a comparison
with results of a recent analysis performed by Nasser et al. (2001).  

\subsection{AM\,CVn}
AM\,CVn is the prototype of the so-called AM\,CVn stars or helium
cataclysmics, a subgroup of the cataclysmic variables. They are thought to
be the end product of the evolution of close binary systems (El-Khoury \&
Wickramasinghe, 2000). The first observations of AM\,CVn have been made by
Malmquist (1936) and Humason \& Zwicky (1947). Greenstein \& Matthews (1957)
classified AM\,CVn as a helium-rich white dwarf, Burbidge et al. (1967) as a
quasi-stellar object, and Wambler (1967) as a hot star. After the discovery
of periodic variability in the light curve (Smak 1967), Warner \& Robinson (1972)
proposed AM\,CVn to be a close binary system with ongoing mass transfer.
Right now, these systems are believed to be interacting white dwarf binary systems,
consisting of a degenerate C/O white dwarf as primary and a semi-degenerate
low-mass secondary, composed of almost pure helium. The secondary fills its Roche
volume and loses mass via Roche-Lobe overflow onto the primary, generating
an accretion disc around it.

AM\,CVn itself has been analysed by Nasser et al. (2001) using TLUSDISC
(Hubeny 1990). They showed that the system consists of a primary of
about 1.1\,$\rm M_\odot$ and a secondary of about 0.09\,$\rm M_\odot$. The
radius of the primary is 4600\,km, and the mass accretion rate is about $3\cdot
10^{-9}\,\rm M_\odot/yr$. For the calculations, we assumed an inner radius of 
1.4\,$\rm{R_\star}$ and an outer radius of 15\,$\rm{R_\star}$, both
values were varied to explore the influence of the disc size onto the
spectrum. The Reynolds number was set to 15\,000, the damping factor
$\zeta$ was chosen to be 0.001. Convective energy transport is not yet
included in {\sc AcDc}, so we had to neglect convection, fortunately without
getting numerical problems in the outer part of the disc.
The number ratio H/He was set to $10^{-5}$, and abundances of carbon,
nitrogen, oxygen, and silicon were assumed to be solar. The self-consistent
inclusion of metals is an improvement over the  
Nasser et al. (2001) models. Some details concerning the ions, levels and lines we
used in our calculations are shown in Table 1. The atomic data are taken
from the opacity project (Seaton et al. 1994) and the Kurucz line lists
(1991). The Lyman, Balmer and Paschen series of H\,{\sc i}, the
Lyman series of He{\sc i} and the Lyman, Balmer, Paschen and
Bracket series of He{\sc ii}, some Lyman and Balmer lines of metals
as well as resonance lines are Stark broadened, all other line profiles are
Doppler broadened. For the line broadening of H we use VCS tables
(Lemke 1997), for the line broadening of He\,{\sc i} we use BCS tables
(Barnard, Cooper \& Shamey 1969) and Griem tables (Griem 1974), for He\,{\sc
  ii} we use VCS tables (Sch\"oning \& Butler 1989).
In total, we calculated 38 individual disc rings.

\begin{table}
\centering
\caption{Some details concerning the ions and the number of NLTE levels and lines used in our calculations.}
\label{tab_atom}
\begin{tabular}{l|r|r||l|r|r} 
Ion          & NLTE levels & lines & Ion           & NLTE levels & lines\\[0.5ex]\hline
H\,{\sc i}   & 16          & 29    &  N\,{\sc ii}  & 2           &  0  \\
H\,{\sc ii}  &  1          & -     &  N\,{\sc iii} & 34          & 67  \\ 
He\,{\sc i}  & 44          & 31    &  N\,{\sc iv}  & 34          & 53  \\
He\,{\sc ii} & 32          & 59    &  N\,{\sc v}   & 36          & 56  \\ 
He\,{\sc iii}&  1          & -     &  N\,{\sc vi}  & 1           &  0  \\ 
C\,{\sc ii}  & 16          & 25    &  O\,{\sc ii}  & 26          & 36  \\ 
C\,{\sc iii} & 58          & 124   &  O\,{\sc iii} & 28          & 37  \\ 
C\,{\sc iv}  & 36          & 86    &  O\,{\sc iv}  & 11          &  5  \\ 
C\,{\sc v}   &  1          &  0    &  O\,{\sc v}   & 6           &  1  \\ 
      \,     &  \,         &  \,   &  O\,{\sc vi}  & 36          & 48  \\
      \,     &  \,         &  \,   &  O\,{\sc vii} & 1           &  0  \\
\end{tabular}
\end{table}

\subsection{Influence of Disc Parameters on the Spectrum}
First, we examined the influence of different inner and outer radii onto the
spectrum of the disc. We varied the inner radius from 1.4 to 2\,$\rm
R_\star$ and the outer radius from 11 to 15\,$\rm R_\star$. 
\begin{figure}
\centering
\includegraphics[width=0.5\textwidth]{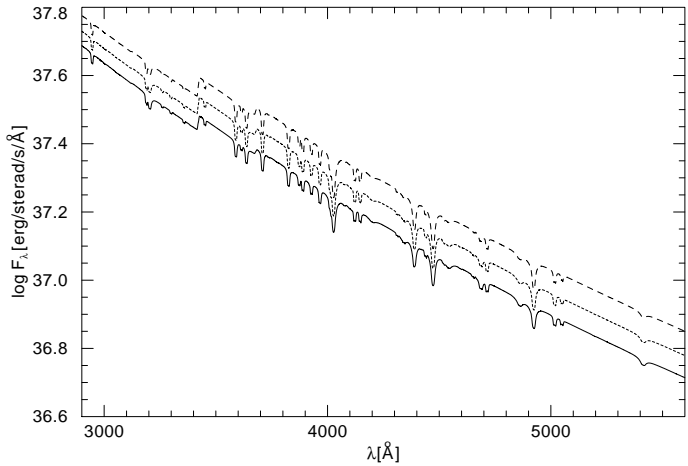}
\caption{Optical spectra of three accretion disc models with outer radii of
  11\,$\rm R_\star$ (solid line), 13\,$\rm R_\star$ (dotted) and
  15\,$\rm R_\star$ (dashed). The mass accretion rate is $3\cdot10^{-9}\,\rm
M_\odot/yr$ and the inclination is $10^\circ$.}
\label{AMCVn_varRad1}
\end{figure}
\begin{figure}
\centering
\includegraphics[width=0.5\textwidth]{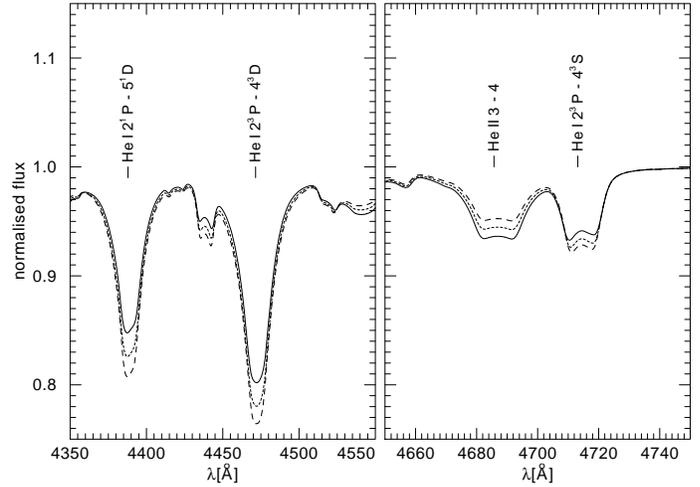}
\caption{Detail of Fig. \ref{AMCVn_varRad1} with three accretion disc models with outer radii
  of 11\,$\rm R_\star$ (solid line), 13\,$\rm R_\star$ (dotted) and
  15\,$\rm R_\star$ (dashed).}
\label{AMCVn_varRad2}
\end{figure}
Figure~\ref{AMCVn_varRad1} shows the optical spectrum of three
disc models with outer radii of 11\,$\rm R_\star$, 13\,$\rm
R_\star$ and 15\,$\rm R_\star$. The mass accretion rate is $3\cdot10^{-9}\,\rm
M_\odot/yr$, the inclination is $10^\circ$. One can clearly see the increasing
total flux with increasing outer radius because of the increasing radiating
surface. Figure \ref{AMCVn_varRad2} shows details of the normalised
spectrum. The line cores of He\,{\sc i} become deeper with increasing
outer radius because the large outer regions of the disc are cool enough to show
strong He\,{\sc i} lines. Variation of the inner radius shows similar, but
smaller effects on the lines.

Second, we analysed the influence of different inclination angles on the
spectrum. 
\begin{figure}
\centering
\includegraphics[width=0.5\textwidth]{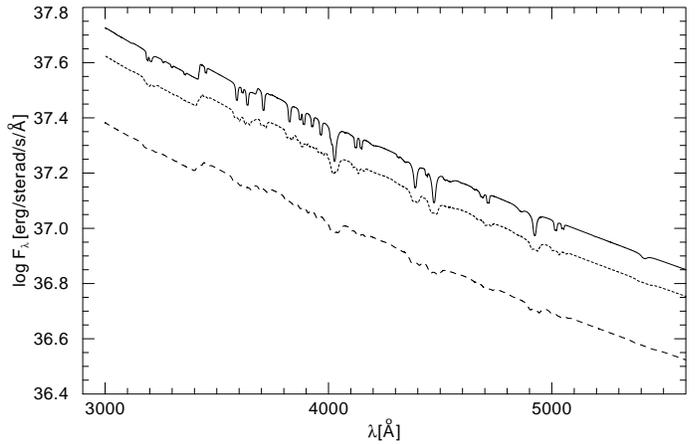}
\caption{Optical spectra of an accretion disc model seen under three different
  inclination angles, 10$^\circ$ (solid line), 36$^\circ$ (dotted) and
  60$^\circ$ (dashed). The inner radius of the disc is 1.4\,$\rm R_\star$, the
  outer radius 15\,$\rm R_\star$ and the mass accretion rate is
  $3\cdot10^{-9}\,\rm M_\odot/yr$.}  
\label{AMCVn_varInk1}
\end{figure}
\begin{figure}
\centering
\includegraphics[width=0.5\textwidth]{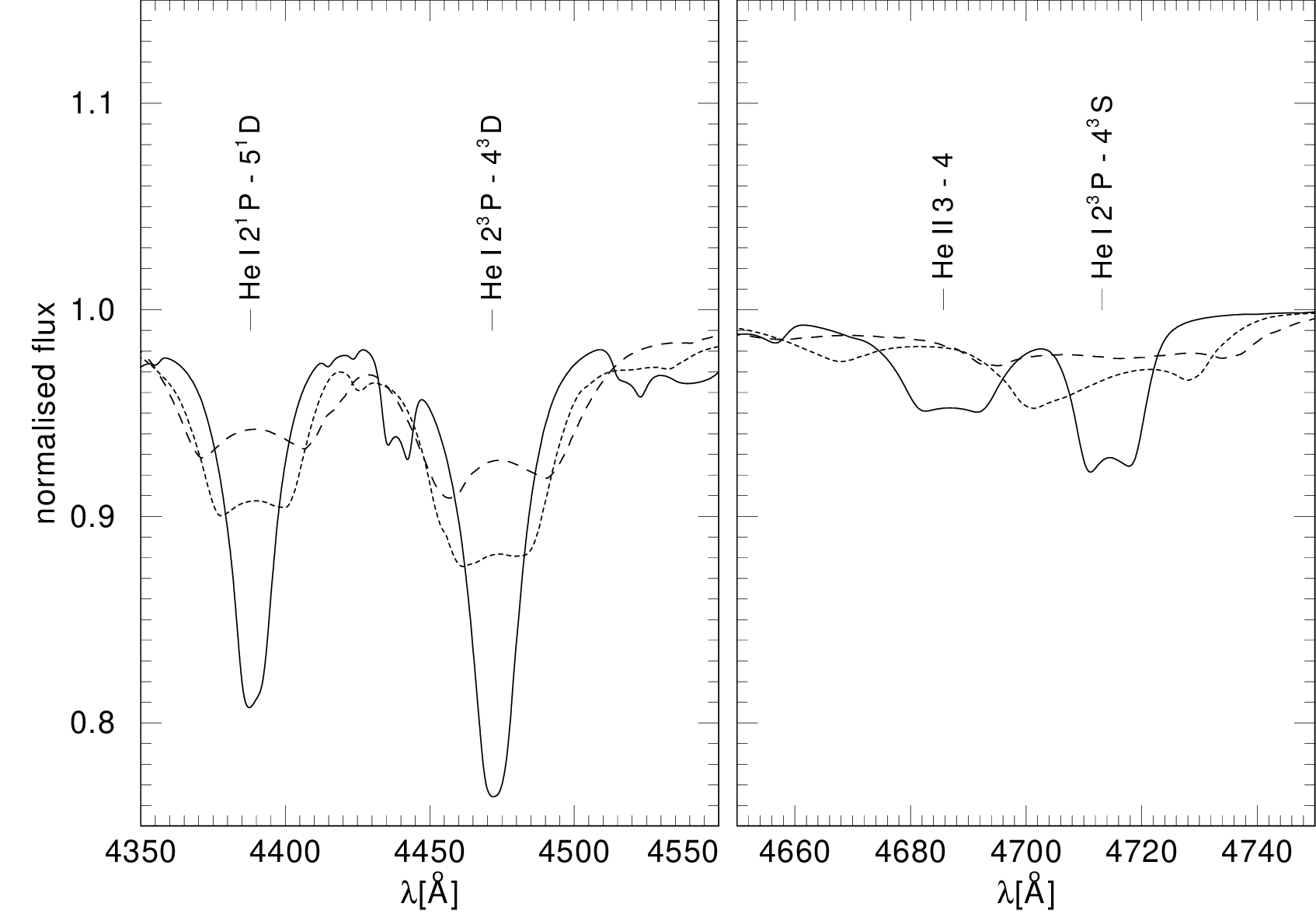}
\caption{Detail of Fig. \ref{AMCVn_varInk1} with an accretion disc model
  seen under three different inclination angles, 10$^\circ$ (solid line),
  36$^\circ$ (dotted) and 60$^\circ$ (dashed).} 
\label{AMCVn_varInk2}
\end{figure}
Figure \ref{AMCVn_varInk1} shows a disc model seen under three different
inclination angles. The total flux decreases with increasing inclination
angle because of the decreasing projected surface.
As shown in Fig. \ref{AMCVn_varInk2}, narrow lines at small inclination angles
become broad lines at high inclination angles because of the increasing
rotational broadening.

\subsection{Vertical Structure}
\begin{figure}
\centering
\includegraphics[height=11cm]{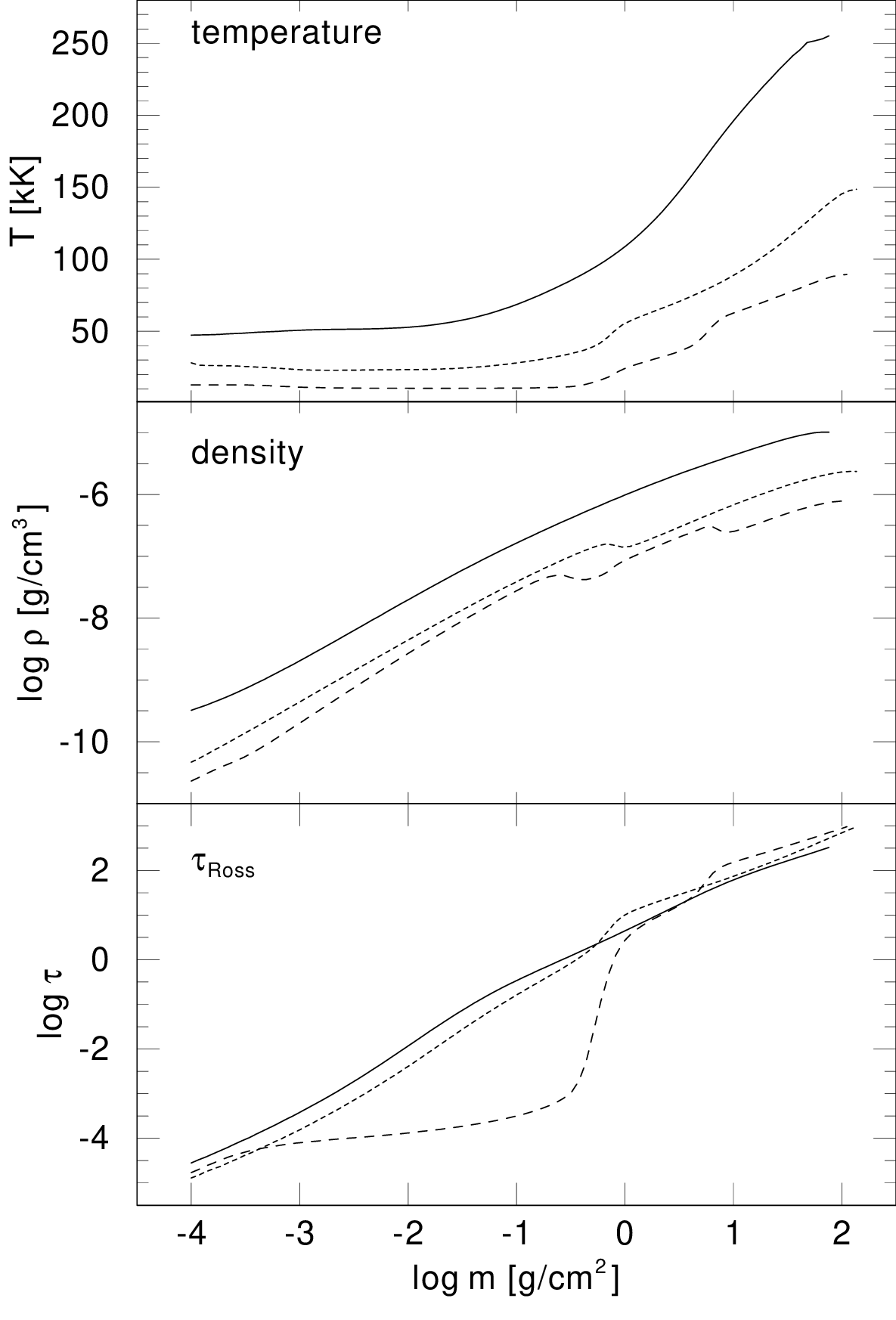}
\caption{Vertical distribution of temperature, density and Rosseland optical
  depth of three representative disc rings with radii of 1.4\,$\rm R_\star$ (solid line), 7\,$\rm
  R_\star$ (dotted line) and 15\,$\rm R_\star$ (dashed line).}
\label{AMCVn_vertikal}
\end{figure}
In Fig. \ref{AMCVn_vertikal} the vertical distribution of temperature,
density and Rosseland optical depth of three representative disc rings at 1.4\,$\rm
R_\star$, 7\,$\rm R_\star$ and 15\,$\rm R_\star$ is shown. According to Eq.
(\ref{tglg}) disc rings from the outer part of the disc are cooler than inner disc
rings. The temperature decreases monotonously from the midplane to the
surface of the rings. Disc rings composed only of H and He, but without C,
N, and O show an increase of the temperature in the outer layers near the surface. 

\subsection{Model vs. Observation}
Finally we compared our accretion disc model spectra with an observed
spectrum of AM\,CVn, obtained at the 6\,m BAT of the SAO in May 2002.

\begin{figure}
\centering
\includegraphics[width=0.5\textwidth]{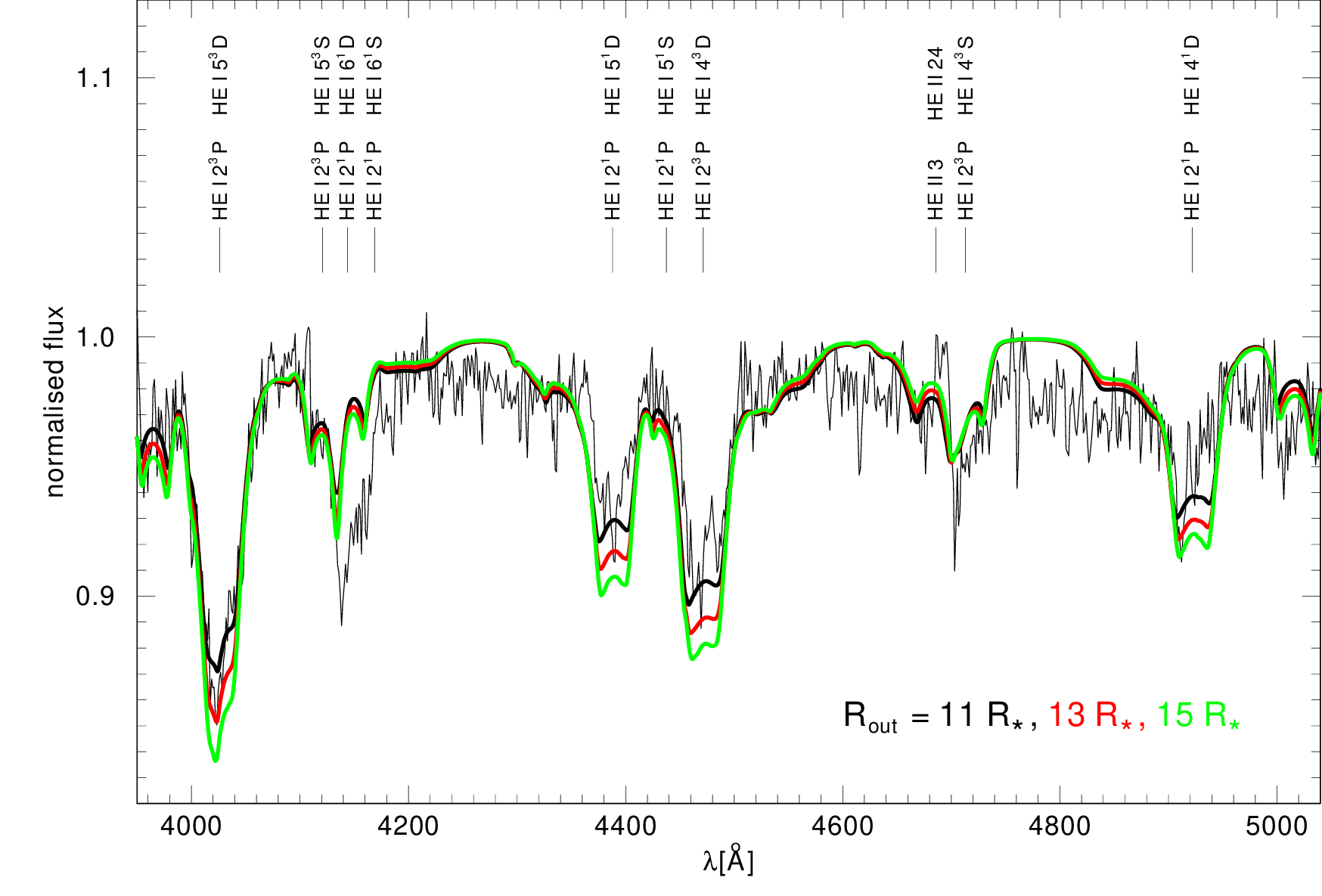}
\caption{Comparison of the observed spectrum (thin line) with model 
  spectra of the AM\,CVn disc for 3 different outer radii.} 
\label{AMCVn_BeobMod1}
\end{figure}

Figure \ref{AMCVn_BeobMod1} shows the observed spectrum and three model
spectra with outer radii of 11\,$\rm R_\star$, 13\,$\rm R_\star$ and
15\,$\rm R_\star$, respectively. The inner radius is 1.4\,$\rm R_\star$, the 
inclination is 36$^\circ$ and the mass accretion rate is $3\cdot10^{-9}\,\rm
M_\odot/yr$. Owing to the larger and at the same time cooler radiating
surface, a larger outer radius leads to an increase of the spectral line
strengths of neutral helium compared to those of ionized helium.

\begin{figure}
\centering
\includegraphics[width=0.5\textwidth]{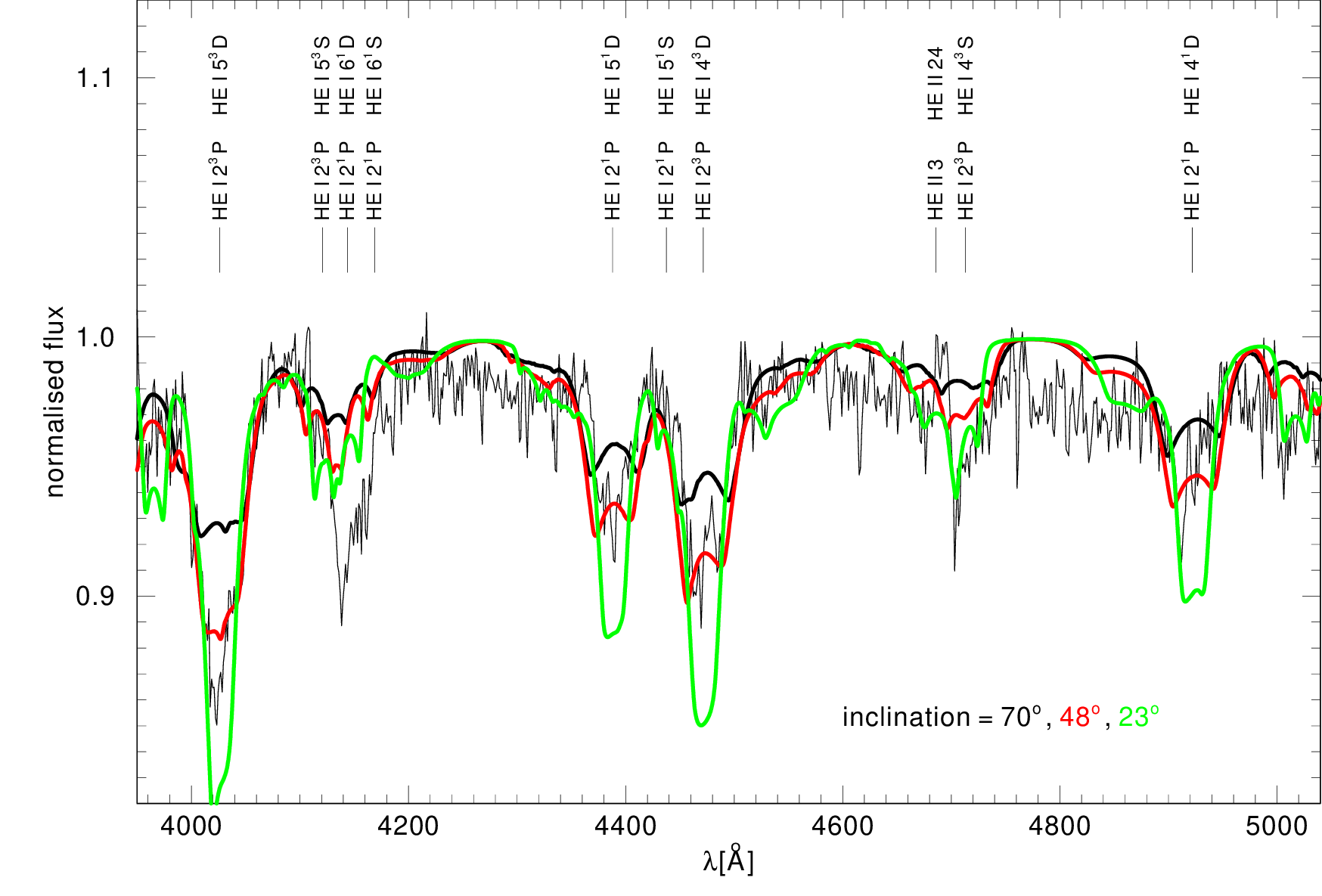}
\caption{Comparison of the observed spectrum (thin line) with model 
  spectra of the AM\,CVn disc for 3 different inclination angles.} 
\label{AMCVn_BeobMod2}
\end{figure}

Figure \ref{AMCVn_BeobMod2} shows the influence of the inclination angle
on the spectrum. The disc models with inclination angles of 23$^\circ$,
  48$^\circ$ and 70$^\circ$ extend from 1.4\,$\rm R_\star$ to
13\,$\rm R_\star$ and the mass accretion rate is $3\cdot10^{-9}\,\rm
M_\odot/yr$. The larger the inclination angle, the stronger the spectral
lines are broadened due to the increasing radial component of the Kepler
velocity. 

\begin{figure}
\centering
\includegraphics[width=0.5\textwidth]{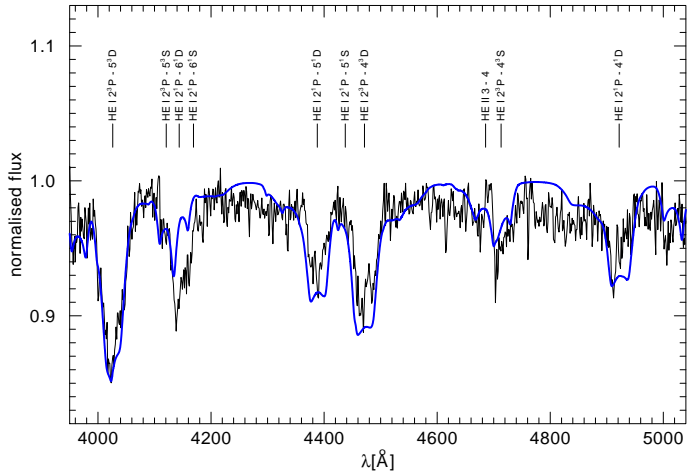}
\caption{Comparison of the observed spectrum (thin line) with a model 
  spectrum of the AM\,CVn disc.} 
\label{AMCVn_BeobModbest}
\end{figure}

In Fig. \ref{AMCVn_BeobModbest} our best fit is shown. The disc model
extends from 1.4\,$\rm R_\star$ to 13\,$\rm R_\star$, the 
inclination is 36$^\circ$ and the mass accretion rate is $3\cdot10^{-9}\,\rm
M_\odot/yr$. The shapes of the He\,{\sc i} lines are in good agreement
with the observation. 

\section{Conclusions}

   \begin{enumerate}
      \item We developed the new code {\sc AcDc} for detailed NLTE calculations of
        accretion disc spectra of cataclysmic variables and compact X-ray
        binaries, solving the radiation transfer equation self-consistently
        together with the structure equations under consideration of full
        metal line blanketing as well as irradiation of the accretion disc
        by the central object. 
      \item Variation of the extension of the disc has a significant influence on
        the spectrum of the AM\,CVn disc. Lines of He\,{\sc i} become deeper
        with increasing outer radius. In addition, the inclination angle of
        the disc has a clear influence on the spectral lines by Kepler rotation.
      \item The comparison of model spectra with an observed spectrum of
        AM\,CVn yields 1.4\,$\rm R_\star$ for the inner radius and 13\,$\rm
        R_\star$ for the outer radius, the inclination is found to be
        36$^\circ$ and the mass accretion rate $3\cdot10^{-9}\,\rm
        M_\odot/yr$. This is in agreement with results of Nasser et al. (2001).
   \end{enumerate}
   For future work, we plan to include further physical processes in the program,
   e.g. Comptonisation as well as convective energy transport, into the
   program. Furthermore, we have started to study winds from the accretion
   disc. A depth dependent atomic model will be employed to overcome
   numerical instabilities in irradiated discs due to extremely weak
   populated atomic levels at some depths. 

\begin{acknowledgements}
      This research was supported by the Deut\-sche
      For\-schungs\-ge\-mein\-schaft, DFG grant We~1312/24-1,2 and by the
      DLR under grant 50\,OR\,0201 (TR). We thank Ivan Hubeny for many
      helpful discussions. 
\end{acknowledgements}

\end{document}